\documentclass{appolb}
\usepackage{epsfig}

\begin{document}
\title{ Re-evaluation of the isoscalar mixing angle within selected mesonic nonets%
\thanks{$^{\dagger}$ fxchao@zzuli.edu.cn}%
}
\author{Xue-Chao Feng $^{1\dagger}$, Ke-Wei Wei$^{2}$
\address{$^{1}$College of Physics and Electronic Engineering, Zhengzhou University of Light Industry, 450002 Zhengzhou, China }
\address{$^{2}$ School of Science,Henan University of Engineering, 451191 Zhengzhou, China}
 } \maketitle
\begin{abstract}
 Based on the relations from the meson-meson mass mixing matrix, the mixing angles of isoscalar state have been re-evaluated via mass relations and latest experimental results.  The results in the present work are compared with the values from different theoretical models, meanwhile, the quarkonia content of isoscalar state are presented. In order to check the validity of analysis, some predictions on the decays of the isoscalar state are presented. These predictions may be useful for the phenomenological analysis for meson nonet in future experiments.
\end{abstract}
\section{Introduction}
Quantum Chromodynamics (QCD) was proposed in the 1970s as the basic theory to describe the hadrons and their strong interactions. However, understanding of the strong interactions is far from complete. One of the open problems is the difficulty to interpret the nature of the experimental data from the first principles. Building models, which capture the most important features of strong QCD, is one way to resolve this problem. Therefore, the spectroscopy and the phenomenological description of conventional mesons become important and a series of theoretical models are built to investigate the meson properties in the hadronic physics\cite{Jafarzade,Polyakov79,Maris297,Godfrey189,Eichten4116}.

In the quark model, conventional mesons are bound states of quarks $q$ and antiquarks $\bar{q'}$ (the flavors of $q$ and $q'$ may be different and the spin is $1/2$ ). The quark and antiquark spins can couple to give a total spin $0$ and $1$. The total spin couples the orbital angular momentum result in the total angular momentum $J$. Therefore, meson parity and charge conjugation are determined by $P=(-1)^{L+1}$ and $C=(-1)^{L+S}$, respectively. The mesons are classified in $J^{PC}$ multiplets, namely, $\rho(770)$ ,$K^{*}(892)$, $\omega(782)$ and $\phi(1020)$ as multiplets of $J^{PC}=1^{--}$, $a_{2}(1320)$ ,$K_{2}^{*}(1430)$, $f_{2}(1270)$ and $f^{'}_{2}(1525)$ as multiplets of $J^{PC}=2^{++}$, $\rho_{3}(1690)$ ,$K_{3}^{*}(1780)$, $\omega_{3}(1670)$ and $\phi_{3}(1850)$ as multiplets of $J^{PC}=3^{--}$, etc. According to the ``Review of Particle Physics'' by the Particle Data Group(PDG) in 2020, the nine $q\bar{q'}$ combinations containing the light quark up, down and strange quarks are grouped into an octet and a singlet of light quark mesons\cite{PDG}: $$3\otimes\bar{3}=8\oplus1$$
Due to the SU(3)-symmetry breaking the isoscalar physical states appear as mixtures of the singlet and octet members. The singlet-octet mixing is also called SU(3) mixing. Considered the mixing can provide clues for testing QCD and the constituent quark model, the study of mixing become the focus of many studies in the literature in the last two decades\cite{Brisudova114015,Feng101101}. In the present work, we re-evaluate the mixing angles and the decays of meson nonet via the latest experimental results. The phenomenological description is meaningful for understanding  the nature of new resonances.
\section{State mixing and determination of the mixing angles}
As discussed above section, mesons are $q\bar{q'}$ bound states of quark $q$
and antiquark $\bar{q'}$ in the quark model. In general, the two bare isoscalar states
can mix, which result in two physical isoscalar states.
In the non-strange $F_{n}\equiv(u\bar{u}+d\bar{d})/\sqrt{2}$ and strange
$F_{s}\equiv s\bar{s}$ basis, the mass-squared matrix describing the mixing of
the two physical isoscalar states can be written
as\cite{Brisudova114015,Feng101101}
\begin{equation}
M^2=\left(\begin{array}{cc}
M^2_{F_{n}}+2A & \sqrt{2}AX \\
\sqrt{2}AX  & M^2_{F_{s}}+AX^{2}
\end{array}\right)
\end{equation}
where $M_{F_{n}}$ and $M_{F_{s}}$ are the masses of bare states $F_{n}$ and $F_{s}$, which is
widely adopted assumption\cite{Kawai262,Feldmann114006,Burakovsky489}. $A$ are the mixing parameter which describe the
$q\bar{q}\leftrightarrow q'\bar{q'}$ transition amplitudes\cite{Giacosa091901,Giacosa05012}. The X is
a phenomenological parameter describes the $SU(3)$ broken ratio of the non-strange and strange quark propagators via the constituent
quark mass ratio. In present work, the constituent quark masses in different phenological models are shown in Table1. In Table1, $m_{n}$ and $m_{s}$ denote the constituent quarks masses(where and below $n$ denotes up and down quark).

\begin{center}
\indent\\ \footnotesize Table1 Constituent quark  masses(in $MeV$)
in different phenological models. \begin{tabular}{llllllllll}
\\ \hline
 $Mass$&$m_n(n=u,d)$       &$m_s$    & $X=\frac{m_{n}}{m_{s}}$  &
\\ \hline
Ref.\cite{Godfrey189}      &220&419&0.53
\\
Ref.\cite{Burakosky251}    &229&460&0.63
\\
Ref.\cite{Karliner0307243} &360&540&0.67
\\
Ref.\cite{Scadron735}      &337.5&486&0.70
\\
Ref.\cite{Jovanovic105011} &311&487&0.64
\\
Ref.\cite{Lavelle1}        &310&483&0.64
\\
Average values             &321&491.2&0.66
\\ \hline
\end{tabular}
\end{center}

In a meson nonet, the isoscalar physical states $\varphi$ and $\varphi'$ are the eigenstates of mass-squared matrix and the masses square of $M^2_{\varphi}$ and $M^2_{\varphi'}$ are the eigenvalues, respectively. In the present work, $\varphi$ is mainly nonstrange component and $\varphi'$ is mainly strange component. The physical states $\varphi$ and $\varphi'$ can be related to the $F_{s}$ and $F_{n}$  by
 \begin{equation}
\left(\begin{array}{c}
|\varphi\rangle\\
|\varphi'\rangle
\end{array}\right)=U\left(\begin{array}{c}
  F_{n}\\
  F_{s}
\end{array}\right)=
\left(\begin{array}{cc}
 \cos\beta   &  \sin\beta \\
 -\sin\beta  &  \cos\beta
\end{array}\right)\left(\begin{array}{c}
  F_{n}\\
  F_{s}
\end{array}\right)
\end{equation}
where $\beta$ is the mixing angle in the basis $F_{s}$ and
$F_{n}$. The unitary matrix $U$ can be described as

\begin{equation}
M^{2}=U^{\dag}\left(\begin{array}{cc}
M^2_{\varphi}&0\\
0&M^2_{\varphi'}
\end{array}\right)U
\end{equation}

In addition to the relation (2), the mix of the physical isoscalar states can be also
described as
\begin{equation}
\left(\begin{array}{c}
|\varphi\rangle\\
|\varphi'\rangle
\end{array}\right)=
\left(\begin{array}{cc}
 \cos\theta   & \sin\theta \\
 -\sin\theta  & \cos\theta
\end{array}\right)\left(\begin{array}{c}
\phi_{1}\\
\phi_{8}
\end{array}\right)
\end{equation}
with $$\phi_{1}=(u\bar{u}+d\bar{d}+s\bar{s})/\sqrt{3} $$
$$\phi_{8}=(u\bar{u}+d\bar{d}-2s\bar{s})/\sqrt{6}$$
where $\theta$ is the SU(3) singlet-octet mixing angle.

With the help of
\begin{equation}
\left(\begin{array}{c}
\phi_{1}\\
\phi_{8}
\end{array}\right)=\left(\begin{array}{cc}
\sqrt{\frac{2}{3}}& \sqrt{\frac{1}{3}}\\
\sqrt{\frac{1}{3}}&-\sqrt{\frac{2}{3}}
\end{array}\right)\left(\begin{array}{c}
F_{n}\\
F_{s}
\end{array}\right)
\end{equation}
The following relation is obtained.
\begin{equation}
\left(\begin{array}{cc}
 \cos\beta   & \sin\beta  \\
 -\sin\beta   & \cos\beta
\end{array}\right)
=\left(\begin{array}{cc}
 \cos\theta   & \sin\theta \\
 -\sin\theta  & \cos\theta
\end{array}\right)\left(\begin{array}{cc}
\sqrt{\frac{2}{3}}& \sqrt{\frac{1}{3}}\\
\sqrt{\frac{1}{3}}& -\sqrt{\frac{2}{3}}
\end{array}\right)
\end{equation}

From the (1), (2) and (6), the following relations are obtained.
\begin{equation}
M^2_{F_{n}}+2A=(\sqrt{\frac{1}{3}}\cos\theta-\sqrt{\frac{2}{3}}\sin\theta)^{2}M^2_{\varphi'}+(\sqrt{\frac{2}{3}}\cos\theta+\sqrt{\frac{1}{3}}\sin\theta)^{2}M^2_{\varphi}
\end{equation}
\begin{equation}
M^2_{F_{s}}+AX^{2}=(\sqrt{\frac{1}{3}}\cos\theta-\sqrt{\frac{2}{3}}\sin\theta)^{2}M^2_{\varphi}
+(\sqrt{\frac{2}{3}}\cos\theta+\sqrt{\frac{1}{3}}\sin\theta)^{2}M^2_{\varphi'}
\end{equation}
\begin{equation}
\sqrt{2}AX=(\sqrt{\frac{1}{3}}\cos\theta-\sqrt{\frac{2}{3}}\sin\theta)(\sqrt{\frac{2}{3}}\cos\theta+\sqrt{\frac{1}{3}}\sin\theta)(M^2_{\varphi}-M^2_{\varphi'})
\end{equation}

One can see from the above relation, the mixing angle and corresponding physical states are correlated. Inserting the masses of corresponding physical states and the constituent quark mass ratio into the relations (7), (8) and (9), we can obtain the SU(3) singlet-octet mixing angle and the results are shown in Table2. In the present work, considering the fact that $F_{n}\equiv(u\bar{u}+d\bar{d})/\sqrt{2}$  is the orthogonal partner of the isovector state of a meson nonet, one can expect that $F_{n}$
degenerates with isovector state $M_{F_{n}}=M_{I=1}$\cite{Kawai262,Feldmann114006,Burakovsky489,Li743}. Here and below, all the masses used as input for our calculation are taken from the PDG (Table3).

\begin{center}
\indent\\ \footnotesize Table2 The SU(3) singlet-octet mixing angle for $1^{3}S_{1}$ , $1^{3}P_{2}$, $1^{3}D_{3}$, $1^{1}S_{0}$ and $2^{1}S_{0}$  meson nonets.

\begin{tabular}{lllllllll}
\\ \hline
 $J^{PC}$,$N^{2S+1}L_{J}$&  $1^{--}$,$1^{3}S_{1}$  & $2^{++}$,$1^{3}P_{2}$ & $3^{--}$,$1^{3}D_{3}$ &$0^{-+}$,$1^{1}S_{0}$ & $0^{-+}$,$2^{1}S_{0}$ &
\\ \hline
$\theta(^{o})$  &$35.9\pm0.1$   &$31.1\pm0.2$     &$32.4\pm0.7$ &$-4.2\pm0.3$  &$34.4\pm13.8$
\\ \hline
\end{tabular}
\end{center}

\begin{center}
\indent\\ \footnotesize Table3 The parameters in the relations (7),(8) and (9).

\begin{tabular}{llllll}
\\ \hline
 $J^{PC}$,$N^{2S+1}L_{J}$&$M_{F_{n}}(MeV)$ & $M_{F_{s}}(MeV) $           &$A(MeV^{2}) $
\\ \hline
$1^{--}$,$1^{3}S_{1}$ &782.66      &1018.09     &0.005799
\\
$2^{++}$,$1^{3}P_{2}$ &1275.5       &1523.84     &0.053521
\\
$3^{--}$,$1^{3}D_{3}$ &1667    &1857.74     &0.035731
\\
$0^{-+}$,$1^{1}S_{0}$ &978.78     &639.02     &0.324512
\\
$0^{-+}$,$2^{1}S_{0}$ &1294     &1477.11    &0.601166
\\ \hline
\end{tabular}
\end{center}

From Table2, we find the SU(3) singlet-octet mixing angle of $1^{3}S_{1}$, $1^{3}P_{2}$ and $1^{3}D_{3}$ are consistent with existing experimental results and other theoretical predictions\cite{PDG,Isgur122,Crystal451,Bramon271}. Moreover, we also present the mixing angle $\eta$ and $\eta'$ in the $1^{1}S_{0}$ meson nonet. The nature of $\eta$ and $\eta'$ meson is a long standing subject in hadron physics, due to it can provide important information of low energy dynamics of QCD. Nonetheless, there is apparent disagreement for the $1^{1}S_{0}$ meson nonet. The reason may be that we have not take into account the mixing with glueball in our calculation.
In the Refs.\cite{Kou054027,Ambrosino267}, authors indicate that the $\eta'$ may be has a large glueball content.

Based on the relations (2) and (4), we obtain the quarkonia contents for the $1^{3}S_{1}$, $1^{3}P_{2}$, $1^{3}D_{3}$, $1^{1}S_{0}$ and $2^{1}S_{0}$ meson nonet in the non-strange $n\bar{n}=(u\bar{u}+d\bar{d})/\sqrt{2}$ and strange $s\bar{s}$ basis. The results are shown in Table4.

\begin{center}
\indent\\ \footnotesize Table4 The quarkonia content of the isoscalar state for the $1^{3}S_{1}$, $1^{3}P_{2}$, $1^{3}D_{3}$, $1^{1}S_{0}$ and $2^{1}S_{0}$ meson nonet

\begin{tabular}{llllllllll}
\\ \hline
 $J^{PC}$,$N^{2S+1}L_{J}$& $\varphi$ & $\varphi'$& $\beta(^{o})$               &$cos\beta$&  $sin\beta$&
\\ \hline
$1^{--}$,$1^{3}S_{1}$&$\omega(782)$ & $\phi(1020)$& $-0.6$           &0.9999&  -0.0105&
\\
$2^{++}$,$1^{3}P_{2}$ &$f_{2}(1270)$ & $f'_{2}(1525)$& $4.2$         &0.9971&  0.0733&
\\
$3^{--}$,$1^{3}D_{3}$ &$\omega_{3}(1670)$ & $\phi_{3}(1850)$& $2.9$  &0.9987&  0.0506&
\\
$0^{-+}$,$1^{1}S_{0}$ &$\eta'(958)$ & $\eta$& $-39.5$                 &0.7716&  -0.6361&
\\
$0^{-+}$,$2^{1}S_{0}$&$\eta(1295)$ & $\eta(1475)$& $-0.9$             &0.9998&  -0.0157&
\\ \hline
\end{tabular}
\end{center}
\section{Decays and mixing}
In order to check the consistence of our results in Table4, we compared the decays of isoscalar of meson nonet with the experimental data. According to Refs.\cite{Bramon271,Li2775}, we have the following relations.

For the $1^{3}S_{1}$ meson state
\begin{equation}
\frac{\Gamma(\phi\rightarrow\pi\gamma)}{\Gamma(\omega\rightarrow\pi\gamma)}=\frac{\sin^{2}\beta_{(1^{3}S_{1})}}{\cos^{2}\beta_{(1^{3}S_{1})}}\left(\frac{(M^2_{\phi}-M^2_{\pi})M_{\omega}}{(M^2_{\omega}-M^2_{\pi})M_{\phi}}\right)^3
\end{equation}
\begin{equation}
\frac{\Gamma(\phi\rightarrow\pi^{+}\pi^{-})}{\Gamma(\omega\rightarrow\pi^{+}\pi^{-})}=\frac{\sin^{2}\beta_{(1^{3}S_{1})}}{\cos^{2}\beta_{(1^{3}S_{1})}}\left(\frac{\sqrt{M^2_{\phi}-4M^2_{\pi^{\pm}}}}{\sqrt{M^2_{\omega}-4M^2_{\pi^{\pm}}}}\right)^3
\end{equation}

For the $1^{3}P_{2}$ meson state
\begin{equation}
\frac{\Gamma(f_{2}(1270)\rightarrow\gamma\gamma)}{\Gamma(a_{2}(1320)\rightarrow\gamma\gamma)}=\frac{1}{9}(5\cos^{2}\beta_{(1^{3}P_{2})}+\sqrt{2}\sin^{2}\beta_{(1^{3}P_{2})})^2\left(\frac{M_{f_{2}(1270)}}{M_{a_{2}(1320)}}\right)^3
\end{equation}

\begin{equation}
\frac{\Gamma(f'_{2}(1525)\rightarrow\gamma\gamma)}{\Gamma(a_{2}(1320)\rightarrow\gamma\gamma)}=\frac{1}{9}(5\sin^{2}\beta_{(1^{3}P_{2})}-\sqrt{2}\cos^{2}\beta_{(1^{3}P_{2})})^2\left(\frac{M_{f'_{2}(1525)}}{M_{a_{2}(1320)}}\right)^3
\end{equation}

For the $2^{1}S_{0}$ meson state
\begin{equation}
\frac{\Gamma(\eta_{(1295)}\rightarrow\gamma\gamma)}{\Gamma(\pi_{(1300)}\rightarrow\gamma\gamma)}=\frac{1}{9}(5\cos^{2}\beta_{(2^{1}S_{0})}+\sqrt{2}\sin^{2}\beta_{(2^{1}S_{0})})^2\left(\frac{M_{\eta_{1295}}}{M_{\pi_{1300}}}\right)^3
\end{equation}

\begin{equation}
\frac{\Gamma(\eta_{(1475)}\rightarrow\gamma\gamma)}{\Gamma(\pi_{(1300)}\rightarrow\gamma\gamma)}=\frac{1}{9}(5\sin^{2}\beta_{(2^{1}S_{0})}-\sqrt{2}\cos^{2}\beta_{(2^{1}S_{0})})^2\left(\frac{M_{\eta_{1475}}}{M_{\pi_{1300}}}\right)^3
\end{equation}

$$
\frac{\Gamma(\eta (1295)\rightarrow\omega\gamma)} {\Gamma(\eta
(1295)\rightarrow\rho\gamma)}=\frac{1}{9}
\left(\frac{M^2_{\eta(1295)}-M^2_\omega}{M^2_{\eta(1295)}-M^2_\rho}\right)^3
$$
\begin{equation}
 \times\left(\frac{\cos\beta_{(1^{3}S_{1})}\cos\beta_{(2^{1}S_{0})}-2\sin\beta_{(1^{3}S_{1})}\sin\beta_{(2^{1}S_{0})}}{\cos\beta_{(2^{1}S_{0})}}\right)^2
\end{equation}

$$
\frac{\Gamma(\eta (1475)\rightarrow\omega\gamma)} {\Gamma(\eta
(1475)\rightarrow\rho\gamma)} =\frac{1}{9}
\left(\frac{M^2_{\eta(1475)}-M^2_\omega}{M^2_{\eta(1475)}-M^2_\rho}\right)^3
$$
\begin{equation}
\times \left(\frac{\cos\beta_{(1^{3}S_{1})}\sin\beta_{(2^{1}S_{0})}+2\sin\beta_{(1^{3}S_{1})}\cos\beta_{(2^{1}S_{0})}}{\sin\beta_{(2^{1}S_{0})}}\right)^2
\end{equation}

$$
\frac{\Gamma(\eta (1295)\rightarrow\phi\gamma)} {\Gamma(\eta
(1295)\rightarrow\rho\gamma)} =\frac{1}{9}
\left(\frac{M^2_{\eta(1295)}-M^2_\phi}{M^2_{\eta(1295)}-M^2_\rho}\right)^3
$$
\begin{equation}
\times \left(\frac{\sin\beta_{(1^{3}S_{1})}\cos\beta_{(2^{1}S_{0})}+2\cos\beta_{(1^{3}S_{1})}\sin\beta_{(2^{1}S_{0})}}{\cos\beta_{(2^{1}S_{0})}}\right)^2
\end{equation}

$$
\frac{\Gamma(\eta (1475)\rightarrow\phi\gamma)} {\Gamma(\eta
(1475)\rightarrow\rho\gamma)} =\frac{1}{9}
\left(\frac{M^2_{\eta(1475)}-M^2_\phi}{M^2_{\eta(1475)}-M^2_\rho}\right)^3
$$
\begin{equation}
\times \left(\frac{\sin\beta_{(1^{3}S_{1})}\sin\beta_{(2^{1}S_{0})}-2\cos\beta_{(1^{3}S_{1})}\cos\beta_{(2^{1}S_{0})}}{\sin\beta_{(2^{1}S_{0})}}\right)^2
\end{equation}

\begin{center}
\indent\\ \footnotesize Table5 The predicted results in our framework (10)-(19).
\begin{tabular}{llllllll}
\\ \hline
   Relation  &This work  &Expt.\cite{PDG}             & Relation   &This work& Expt.\cite{PDG}&
\\ \hline
  (10)       &$4.23\times10^{-4}$   &  &(11)        & $3.48\times10^{-4}$   &  $0.0024\pm0.0006$
\\
(12)         &$2.60$                &  $3.03\pm0.40$&(13)        & $0.19$                &$0.0081\pm0.0011$

\\
(14)         &$2.76$                &  &(15)        & $0.29$                &
\\
(16)         &$0.11$                &  &(17)        & $0.59$   &
\\
(18)         &$1.88\times10^{-4}$   &  &(19)        & $1.76\times10^{3}$    &
\\ \hline
\end{tabular}
\end{center}
The predicted results of (10)-(19) are determined as shown in Table 5.

\section{Conclusions}
In summary, we have studied the mixing angles of isoscalar state based on the relations from the meson-meson mass mixing matrix. In order to check the consistence of our results, we compared the values given in different references. On the one hand, from Table2, we can see that the mixing angle of $1^{3}S_{1}$, $1^{3}P_{2}$, and $1^{3}D_{1}$ meson nonet in this work are in agreement with Refs.\cite{PDG,Isgur122}. Meanwhile the quarkonia content and decays of the isoscalar state of meson nonet are presented.
On the other hand, we also find the calculated results for pseudoscalar meson is inconsistent with values from other theoretical models.
The reason may be that we didn't consider the mixing between $\eta$ and $\eta'$ with the pseudoscalar glueball. In the past forty years, the quarkonia-glueball structure of $\eta$ and $\eta'$ has been discussed many times \cite{Kawai262,Frank451,Carvalho95}. Therefore we speculate that the $\eta'$ may be has a large glueball content.

In our work, we didn't discuss the mixing angles between $f_{1}(1285)$ and $f_{1}(1420)$, and between $h_{1}(1170)$ and $h_{1}(1415)$. These two mixing angles are related to the mixing between $K_{1A}$ and $K_{1B}$. The$K_{1A}$ and $K_{1B}$ mixing is investigated in Refs.\cite{Divotgey135,Cheng116,Burakovsky1368,Burakovsky2879} . Although we have not calculated these parameters now, with the further enrichment of experimental data, we can still do a lot of analysis by using the relationship in the future. In addition, if the mixing angle can be determined, the relations we get can also be used to analyze the mass of some physical states.

\textbf{Acknowledgments}
This project supported by the Zhengzhou University of Light Industry Foundation of China (Grant Nos.2009XJJ011 and 2012XJJ008) and the Key Project of Scientific and Technological Research of the Education Department of Henan Province (Grant No 13B140332).


\begin{thebibliography}{00}
\bibitem{Jafarzade} S. Jafarzade, A. Koenigstein and F. Giacosa,{\it Phys.Rev. } {\bf D103}, 096027 (2021).

\bibitem{Polyakov79} A. M. Polyakov,{\it Phys. Lett.} {\bf B59}, 79 (1975).

\bibitem{Maris297} P. Maris and C. D. Roberts, {\it Int. J. Mod. Phys.} {\bf E12}, 297 (2003).

\bibitem{Godfrey189} S. Godfrey and N. Isgur, {\it Phys. Rev.} {\bf D32}, 189 (1985).

\bibitem{Eichten4116}E. J. Eichten, C. T. Hill and C. Quigg, {\it Phys. Rev. Lett.} {\bf 71}, 4116 (1993).

\bibitem{PDG}
   P. A. Zyla et al. [Particle Data Group], {\it Prog. Theor. Exp. Phys.} {\bf 8}, 083C01 (2020).

\bibitem{Brisudova114015}M. M. Brisudova, L. Burakovsky and T. Goldman, {\it Phys. Rev.} {\bf D58}, 114015 (1998).

\bibitem{Feng101101}X. C. Feng and  K. W. Wei, {\it Chin. Phys. Lett.} {\bf 29}, 101101 (2012).

\bibitem{Kawai262} E. Kawai, {\it Phys. Lett.} {\bf B124}, 262 (1983).

\bibitem{Feldmann114006}T. Feldmann, P. Kroll P and B. Stech, {\it Phys. Rev.} {\bf D58}, 114006 (1998).

\bibitem{Burakovsky489} L. Burakovsky and P. R. Page, {\it Eur. Phys. J.} {\bf C12}, 489 (2000).

\bibitem{Giacosa091901} F. Giacosa, A. Koenigstein, and R. D. Pisarski, {\it Phys. Rev.} {\bf D97}, 091901 (2018).


\bibitem{Giacosa05012} F. Giacosa {\it EPJ Web Conf.} {\bf 199}, 05012 (2019).

\bibitem{Burakosky251} L. Burakosky and T. Goldman, {\it Phys. Lett.} {\bf B434}, 251 (1998).

\bibitem{Karliner0307243}M. Karliner and H. J. Lipkin, {\it hep-ph/0307243}

\bibitem{Scadron735} M. D. Scadron, R. Delbourgo and G. Rupp, {\it J. Phys.} {\bf G32}, 735 (2006).

\bibitem{Jovanovic105011} V. B. Jovanovic, {\it Phys. Rev.} {\bf D76}, 105011 (2007).

\bibitem{Lavelle1} M. Lavelle and D. McMullan, {\it Phys. Rept.} {\bf279}, 1 (1997).


\bibitem{Feldmann114006}T. Feldmann, P. Kroll and B. Stech, {\it Phys. Rev.} {\bf D58}, 114006 (1998).

\bibitem{Burakovsky489} L. Burakovsky and P. R. Page,  {\it Euro. Phys. J.} {\bf C12}, 489 (2000).

\bibitem{Li743} D. M. Li, Z. Q. Shi and H. Yu,{\it Mod. phys. Lett.} {\bf A21}, 743 (2006).


\bibitem{Isgur122} N. Isgur, {\it Phys. Rev.} {\bf D13}, 122 (1976).

\bibitem{Crystal451}C. Amsler et al. (Crystal Barrel Collaboration),  {\it Phys. Lett.} {\bf B294}, 451 (1992).

\bibitem{Bramon271}A. Bramon, R. Escribano and M. D. Scadron, {\it Eur. Phys. J.} {\bf C7}, 271 (1999).

\bibitem{Kou054027}E. Kou, {\it Phys. Rev.} {\bf D63}, 054027 (2001).

\bibitem{Ambrosino267}F. Ambrosino et al. (KLOE Collaboration),{\it Phys. Lett.} {\bf B648}, 247 (2007).

\bibitem{Li2775}D. M. Li, B. Ma, Q. K. Yao, J. L. Feng, X. C. Feng and H. Yu, {\it Mod. Phys. Lett.} {\bf A18}, 2775 (2003).

\bibitem{Frank451}M. Frank, P. O'Donell,{\it Phys. Lett.} {\bf B144}, 451 (1984).

\bibitem{Carvalho95}W.S. Carvalho, A.C.B. Antunes, A.S. de Castro,{\it Eur. Phys. J.} {\bf C7}, 95 (1999).

\bibitem{Divotgey135}F. Divotgey, L. Olbrich and F. Giacosa,{\it Eur. Phys. J.} {\bf A49}, 135 (2013).

\bibitem{Cheng116}H. Y. Cheng,{\it Phys. Lett.} {\bf B707}, 116 (2012).

\bibitem{Burakovsky1368}L. Burakovsky and J. T. Goldman, {\it Phys. Rev.} {\bf D56}, R1368 (1997).

\bibitem{Burakovsky2879}L. Burakovsky and J. T. Goldman, {\it Phys. Rev.} {\bf D57}, 2879 (1998).


\end{thebibliography}
\end{document}